# Validation of 4D Monte Carlo dose calculations using a programmable deformable lung phantom

**Sara Gholampourkashi[1], Joanna E. Cylger[1,2], Bernie Lavigne[2] and Emily Heath[1]**

[1] Department of Physics, Carleton University, Ottawa, ON, Canada

[2] Department of Medical Physics, The Ottawa Hospital Cancer Center, Ottawa, ON, Canada

**Abstract**

We present the validation of 4D Monte Carlo (4DMC) simulations to calculate dose deliveries to a deforming anatomy in the presence of realistic respiratory motion traces. A previously developed deformable lung phantom comprising an elastic tumor was modified to enable programming of arbitrary motion profiles. The phantom moving with irregular breathing patterns was irradiated using static and VMAT beam deliveries. Using the RADPOS 4D dosimetry system, point doses were measured inside and outside the tumor. Film was used to acquire dose profiles along the motion path of the tumor (S-I). In addition to dose measurements, RADPOS was used to record the motion of the tumor during dose deliveries. Dose measurements were then compared against 4DMC simulations with EGSnrc/4DdefDOSXYZnrc using the recorded tumor motion. The agreements between dose profiles from measurements and simulations were determined to be within 2%/2 mm. Point dose agreements were within 2σ of experimental and/or positional/dose reading uncertainties. 4DMC simulations were shown to accurately predict the sensitivity of delivered dose to the starting phase of breathing motions. We have demonstrated that our 4DMC method, in combination with RADPOS, can accurately simulate realistic dose deliveries to a deforming anatomy moving with realistic breathing traces. This 4DMC tool has the potential to be used as a quality assurance tool to verify treatments involving respiratory motion. Adaptive treatment delivery might be another area that may benefit from the potential of this 4DMC tool.

## 1. Introduction

4D dose calculation methods account for the effect of respiratory motion on the delivered dose by taking into account all three impacts of respiratory motion, including dose blurring, dose deformations and interplay effects (Brock *et al* 2003, Keall *et al* 2004). In a 4D dose calculation algorithm, dose is calculated on multiple respiratory states of the anatomy and mapped to a reference anatomy by use of deformation vector fields (DVFs) to yield cumulative dose distributions. Different dose mapping algorithms such as dose interpolation mapping (DIM) (Rosu *et al* 2005), energy mass congruent mapping (EMCM) (Zhong and Siebers 2009, Siebers and Zhong 2008) and the voxel warping method (VWM) (Heath and Seuntjens 2006) have been developed.

Prior to clinical implementation, experimental verification of 4D dose calculation algorithms with phantoms that simulate both motion and deformation of different organs (e.g. lung, liver, etc.) as well as the tumor is required. An essential design consideration for such phantoms is their capability to accommodate different dosimeters (e.g. film, ion chamber, MOSFET) to measure the dose delivered during treatment. Reproducibility of the phantom motion and geometry from one setup to the next is another important requirement. Several studies using phantoms to verify 4D dose calculation algorithms have been published (Vinogradskiy *et al* 2009a, 2009b, Niu *et al* 2012, Belec and Clark 2013, Ravkilde *et al* 2014, Zhong *et al* 2016). However, they were limited in the sense that they ignored anatomy deformations or the interplay effect as well as the impact of irregular respiratory motion patterns on the delivered dose. Furthermore, in most of the studies only point dose measurements were possible.

In a previous study (Gholampourkashi *et al* 2018b) we presented experimental evaluation of a 4D Monte Carlo (MC) tool (4DdefDOSXYZnrc) using a novel deformable lung phantom. Our 4DMC tool is capable of simulating continuous motion and deformation of the anatomy using the voxel warping method to accumulate dose deposited in different anatomical states. The method has been described in a previous publication (Gholampourkashi *et al* 2017). Currently, the anatomical motion is modeled using one set of deformation vectors (from end-of-exhale to end-of-inhale) and a measured motion trace is used to scale the deformation vectors to reproduce the realized anatomical states. Previous validation work compared measured and simulated doses during static and VMAT treatment deliveries on an Elekta Infinity linac in the presence of sinusoidal motion. It was found that 4DMC dose calculations agreed within 5%, or better, with the measurements. The present study expands on our previous work to study the impact of irregular respiratory motion patterns on the dose delivered to the phantom. The phantom was further modified to be capable of moving with irregular and patient-derived respiratory profiles. We characterize the phantom motion for different motion patterns and evaluate the accuracy of the programmed motions. Measurements and simulations are compared to assess the accuracy of 4DMC tool to simulate these realistic motion profiles.

## 2. Materials and methods

### 2.1. Deformable programmable lung phantom

#### 2.1.1. Phantom design

Design details of the lung phantom used in this study have been previously published (Gholampourkashi *et al* 2018b). The phantom is made of tissue-equivalent foam and holds a non-rigid tumor inside a cylindrical plug. 32 Lucite beads were injected throughout the phantom to help with image registration as well as quantifying the target registration errors. The previous version of the phantom utilized a DC motor to produce a single amplitude/multi frequency sinusoidal motion using a piston attached to the motor. In this work, the DC motor was replaced with a programmable servo motor to enable simulation of realistic respiratory motion profiles. A picture of the phantom with its components are shown in Figure 1(a). The rotational motion of the motor was converted to linear motion of the piston through a Scotch Yoke mechanism as shown in Figure 1(b).

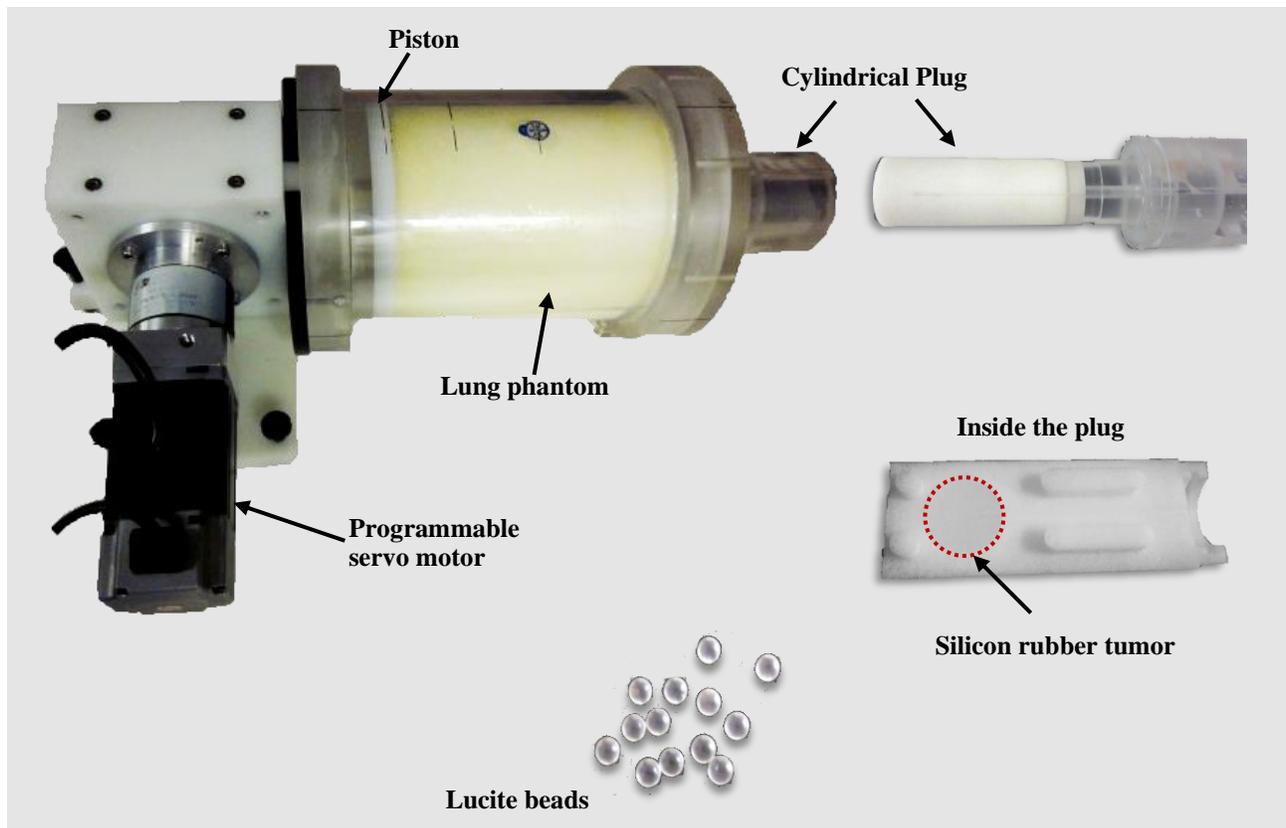

(a)

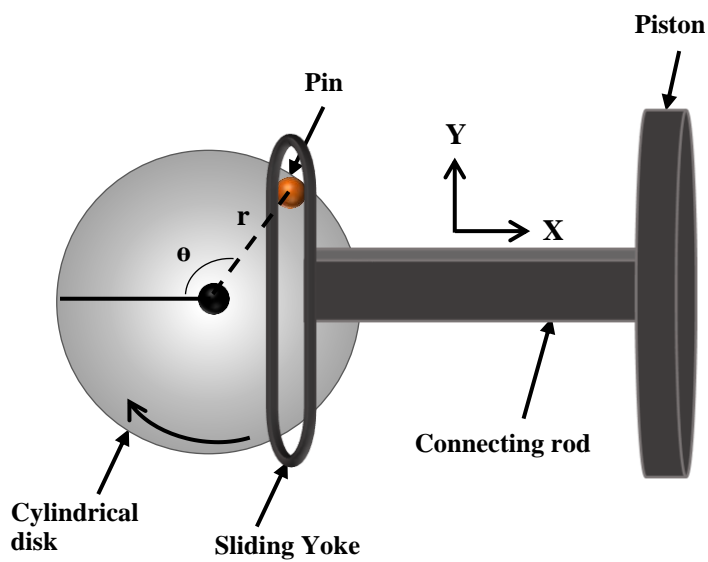

(b)

**Figure 1.** (a) Deformable lung phantom with programmable servo motor to enable simulation of realistic respiratory motion profiles. A cylindrical plug holds the silicon rubber tumor that moves inside the phantom. Details about other components of the phantom are presented elsewhere (Gholampourkashi *et al* 2018b), (b) Diagram of the Scotch Yoke mechanism to convert the rotational motion of the motor to linear motion at the piston. Rotation of the motor disk moves the pin inside the sliding yoke in the vertical (Y) directions and as a result the connecting rod moves in the horizontal (X) direction. The radius of the disk and angle of rotation (i.e. r and θ) determine the amplitude of the linear motion in the X direction.

A cylindrical disk with radius r is attached to the motor disk. Rotation of the motor causes the pin at the edge of the disk to slide in the vertical direction (upward or downward) inside the sliding yoke. This vertical motion results in the horizontal motion of the connecting rod and as a result piston. The magnitude of this linear motion is related to r and θ, which are radius and angle of rotation, respectively and can be calculated by equation 1:

$$X = r\ (1 - \cos\theta) \qquad (1)$$

where r = 2.5 cm and $0° \leq \theta \leq 180°$ for the phantom used in this study. The maximum achievable peak-to-peak (P-P) amplitude at the piston is 5 cm.

The motor can be programmed with wide range of motion profiles, from sinusoidal (varying amplitudes and frequencies) to highly irregular patient respiratory traces. The maximum breathing frequency of the phantom is 4 Hz which correlates to 4 breaths per second or a breathing period of 0.25 s.

### 2.1.2. Motion assessment and reproducibility

In order to validate the motion of the phantom, RADPOS detectors were placed at the tumor center as well as on the top and bottom surfaces of the plug. The detector on the bottom surface of the plug was aligned with the one inside the tumor (i.e. approximately 9 cm from the piston) while the top surface RADPOS was mounted with an offset of 1 cm, towards the inferior side of the phantom, from the other two. This setup is shown in Figure 2 for all three detectors.

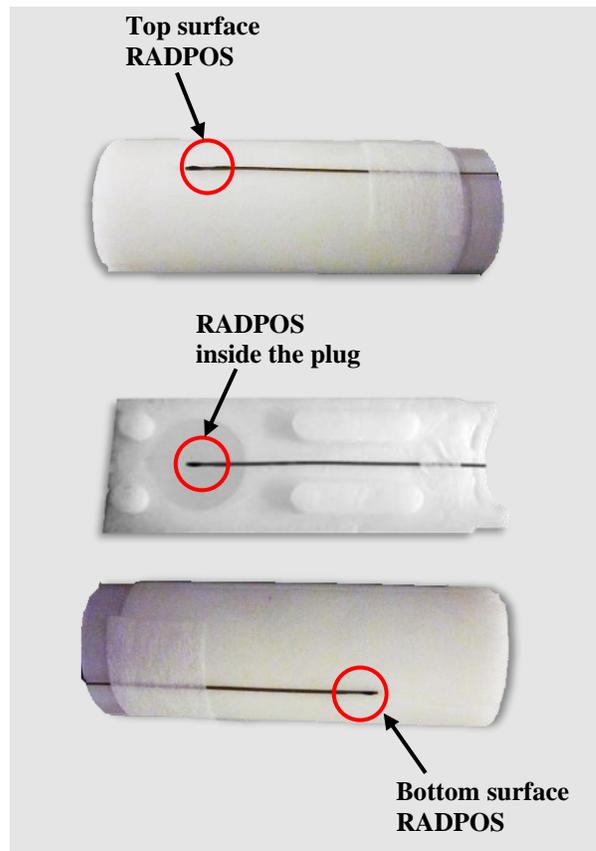

**Figure 2.** RADPOS detectors placed on the top surface (top), inside (middle) and bottom surface (bottom) of the plug to assess the motion of the phantom.

The phantom diaphragm was driven with sinusoidal and irregular motion profiles and the realized motion was recorded in 3D (S-I, A-P and L-R) with a temporal resolution of 100 ms by all three RADPOS detectors. During the measurements, the motion of the diaphragm was recorded by the motion controller for comparison against the motion trace recorded by RADPOS.

Initially, sinusoidal motion profiles with P-P diaphragm amplitudes of 1, 1.5, 2, 2.5 and 3 cm and periods of 2 -7 s in steps of 1 s were tested. Measurements were repeated 6 times over 3 days (i.e. 2 datasets per day) to evaluate both inter-day and intra-day reproducibility and variations of phantom motion. The order in which motion profiles were tested was chosen randomly.

In addition to the sinusoidal motion profiles, 16 different irregular motion traces with P-P diaphragm amplitudes of 1, 1.5, 2, 2.5 and 3 cm were tested to fully evaluate the performance of the phantom. Measurements were repeated 5 times for each motion profile over a 5-day period (i.e. 1 dataset per day). A Python code, which detected the peaks and valleys of the recorded motion traces, was used to compute the average and standard deviation of the P-P motion amplitude for irregular motion patterns.

*2.2. 3DCT acquisition and image registration*

Two sets of 3DCT images of the phantom, in uncompressed and compressed states, were acquired using a helical CT scanner (Brilliance CT Big Bore, Philips, Amsterdam, the Netherlands). The diaphragm displacement for the compressed state was 3 cm, which corresponds to the maximum P-P amplitude in the tested motion profiles. These states correspond to end-of-inhale (EOI) and end-of-exhale (EOE) states of the phantom, respectively. The resulting resolution and matrix size of the 3DCT images were $0.05 \times 0.05 \times 0.2$ cm$^3$ and $512 \times 512 \times 184$. Pitch values and the gantry rotation times for the scans were 0.938 and 0.75 s, respectively.

A deformation vector field (DVF) describing the phantom motion from EOE to EOI breathing phase was generated by registering the respective CT scans in Velocity AI 3.2.0. The structure guided deformable registration feature was used where the position of the Lucite beads in the phantom were used to guide the registration.

*2.3. Treatment plans*

Static and VMAT treatment plans for 6 MV photon beams from an Elekta Infinity linac (Elekta AB., Stockholm, Sweden) were created on the EOI scans of the phantom in Monaco V.5.11.01 (Elekta AB., Stockholm, Sweden). Both plans were aimed to deliver 100 cGy to the center of the tumor which was contoured as the gross tumour volume (GTV) with no margins added to compensate for motion.

The static treatment plan consisted of a single $3 \times 3$ cm$^2$ field. The VMAT plan consisted of 64 control points that delivered a full arc starting and ending at 180° with average angular spacing equal to 5.6°. The corresponding Monaco dose distributions are shown in Figure 3 separately for each treatment plan. The GTV was covered by the 80% and 90% isodose lines on the static and VMAT plans, respectively. The XVMC (X-ray Voxel Monte Carlo) (Fippel 1999) dose calculation algorithm in Monaco was used for dose calculations. Also, in order to be consistent with measurements (film and RADPOS calibrated in Solid Water), dose to water ($D_w$) was calculated in Monaco using a 2 mm dose calculation grid to achieve a statistical uncertainty of 1.0%.

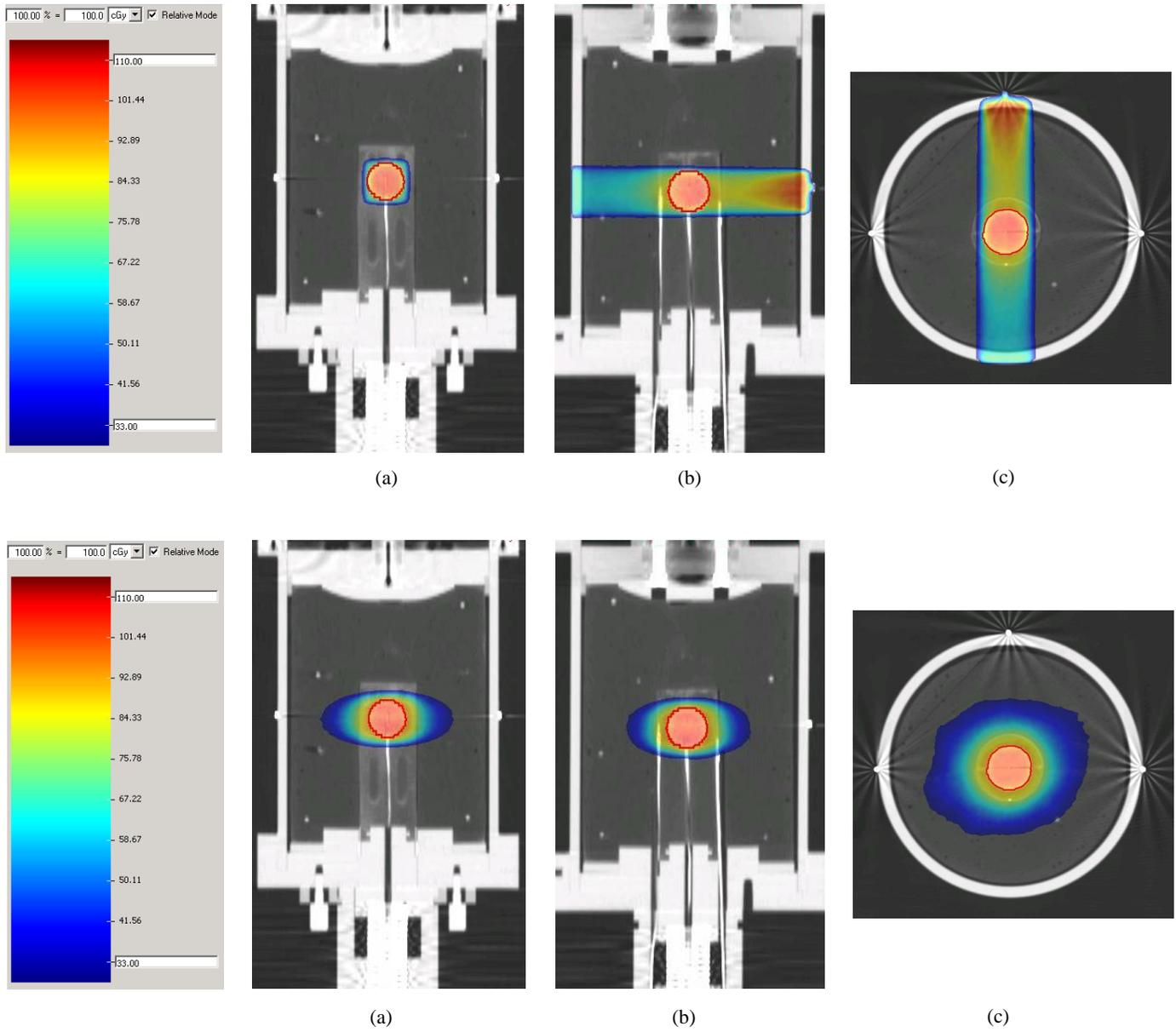

**Figure 3.** Static 3×3 cm² square field (top) and VMAT (bottom) plans: dose distribution from Monaco on (a) coronal, (b) sagittal and (c) axial planes.

## 2.4. Irradiations of the deformable programmable phantom

After exporting treatment plans into the Elekta MOSAIQ RadOnc system, deliveries to the phantom were performed on an Elekta Infinity linac with Agility MLC. The static 3×3 cm² field plan delivered 110.4 MU at a nominal dose rate of 600 MU/min. The VMAT plan delivered 115.9 MU with a varying dose rate. All machine delivery information such as the position of the MLC, jaws, gantry as well as cumulative MU were recorded through delivery log files (IAN V.2; Elekta AB., Stockholm, Sweden) at a temporal resolution of 40 ms.

Three different respiratory motion profiles (Figure 4) were simulated for a P-P diaphragm amplitude of 3 cm and deliveries were repeated three times for each treatment plan. For visibility purposes, only 80 s of the traces are shown here. The respiratory motion profiles shown in Figure 4 included one typical trace (Figure 4(a)), one with large motion variations (Figure 4(b)) and one which resulted in large hysteresis between the motion of the tumor and diaphragm (Figure 4(c)). To study the sensitivity of the delivered dose to the starting phase of the motion, the three deliveries for the first (Figure 4(a)) and second (Figure 4(b)) respiratory profiles were repeated with the beam turned on at approximately 0, 40 and 90 s after starting the phantom motion. The motion traces were recorded at a temporal resolution of 100 ms during deliveries using the RADPOS system. The motion recorded by RADPOS in the S-I direction at the tumor center was compared against the diaphragm motion in Figure 4. The time synchronization between the beam-on and phantom motion were accomplished by synchronizing clock times of the RADPOS and linac computers with the network time protocol (NTP).

For all irradiations the phantom was placed on the couch so that the diaphragm was on the superior side (Figure 5(a)) and the center of the tumor (i.e. plan isocenter) was aligned with the beam isocenter. Point doses inside and outside the plug were measured by calibrated RADPOS detectors that were fixed into special grooves. These grooves were engraved during the molding process of the plug. Calibrated Gafchromic film strips (EBT3, Ashland, Wayne, NJ, USA) were taped on top of the RADPOS probe inside the plug to measure the dose profile in the S-I direction. Film and RADPOS inside the plug are shown in Figure 5(b).

The total dosimetric uncertainties of film and RADPOS detectors measurements were determined to be 2.3%, 2.2% (top RADPOS) and 2.4% (center and bottom RADPOS), respectively (Gholampourkashi *et al* 2017, Cherpak *et al* 2009). The main contributors to these uncertainties were the uncertainties in beam delivery (e.g. depth and field size settings), beam dosimetry calibration (e.g. $N_{D,w}$, $K_Q$)" and reproducibility in ion chamber and RADPOS readings.

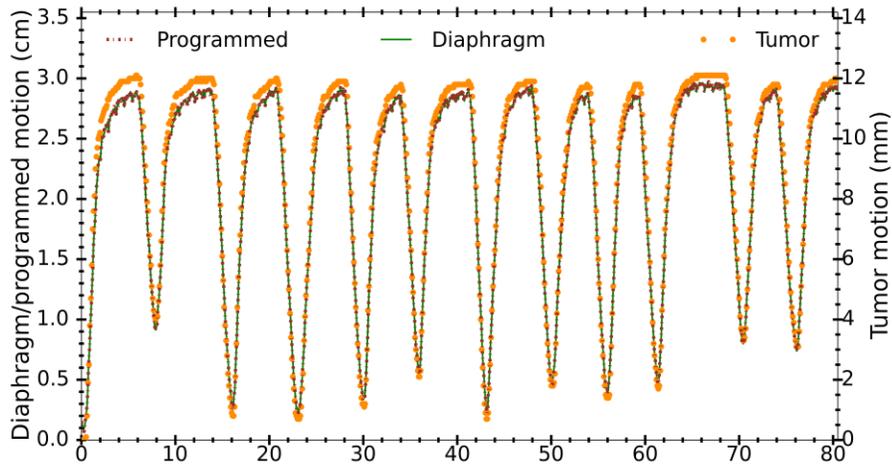

(a)

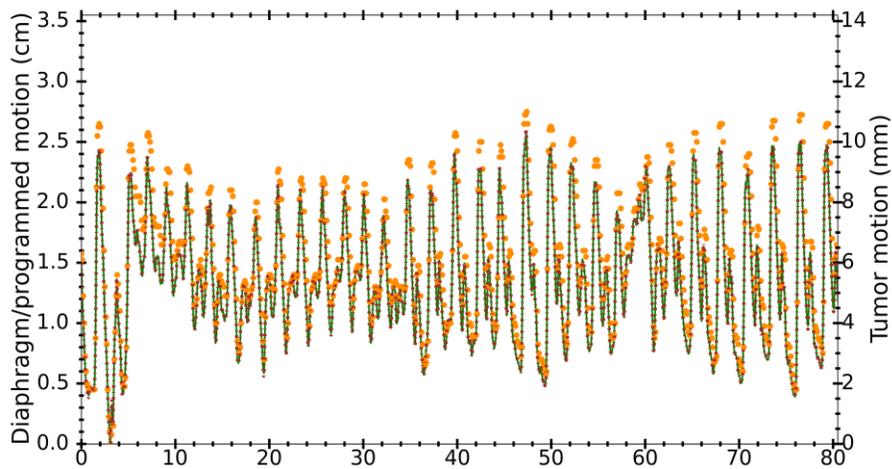

(b)

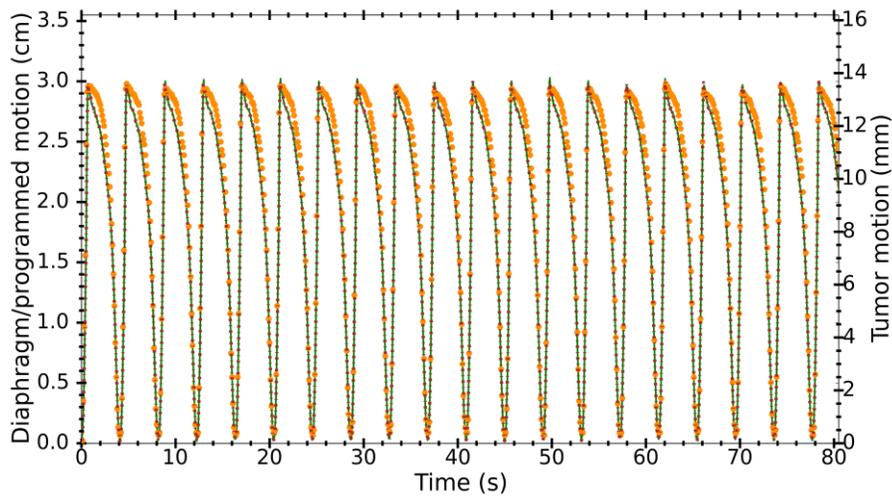

(c)

**Figure 4.** Comparison of the programmed (brown dashed-dotted line), diaphragm (green line) and tumor (orange dots) motion traces for respiratory motion profiles that are (a) typical, (b) very irregular and (c) show large hysteresis between the tumor and diaphragm motion. The P-P diaphragm amplitude was programmed to be 3 cm.

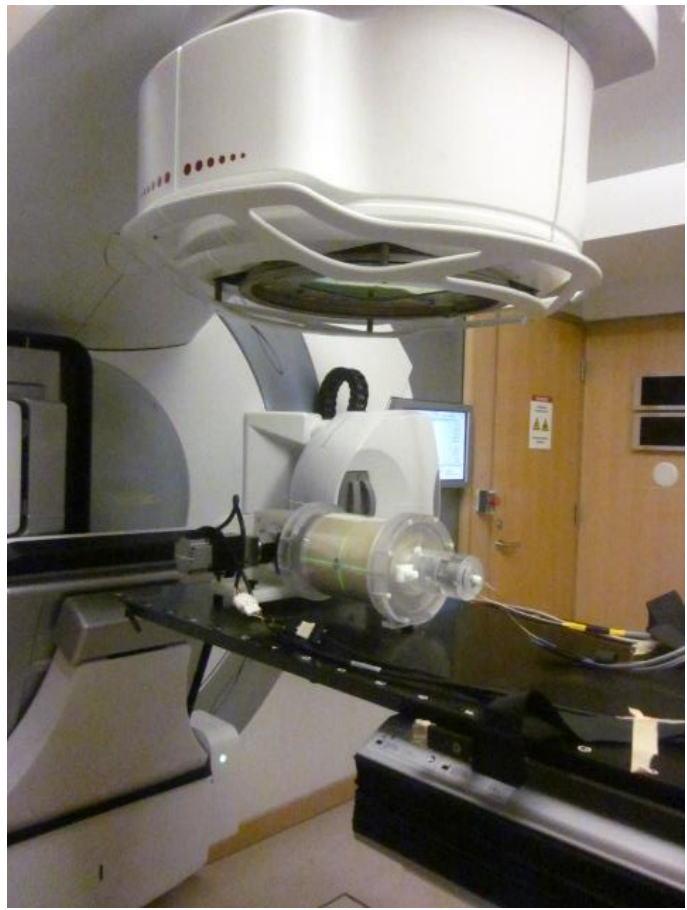

(a)

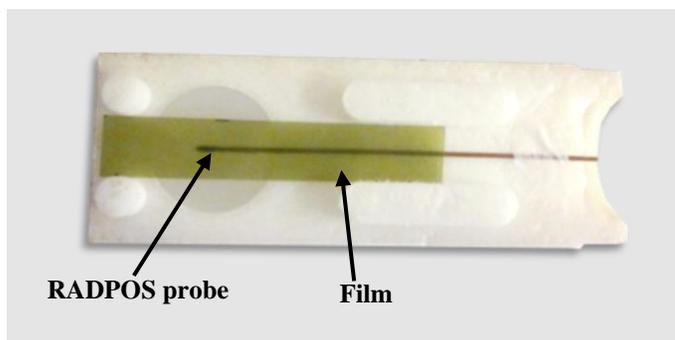

(b)

**Figure 5.** (a) Setup for phantom irradiations such that the piston was in the superior side of the couch and center of the tumor inside the phantom was aligned with the beam isocenter, (b) Film

and RADPOS inside the plug to measure point dose (tumor center) and dose profile along the SI direction. RADPOS is fixed inside an embedded groove and film is taped on top of RADPOS.

*2.5. Monte Carlo simulations*

*2.5.1. User codes and simulation parameters*

EGSnrc (Kawrakow 2013) (V4-2.4.0, National Research Council of Canada, Ottawa, ON, Canada) was used for all simulations in this study. A model of our Elekta Infinity linac with Agility MLC was built using the BEAMnrc (Rogers *et al* 1995) user code and the incident electron beam parameters were tuned according to dose profiles in water (Gholampourkashi *et al* 2018a). The DOSXYZnrc (Walters *et al.*, 2016) and 4DdefDOSXYZnrc (Gholampourkashi *et al* 2017) user codes were used to calculate the resultant dose from MC simulations of the stationary and breathing states of the phantom, respectively. Calculated dose was then converted to absolute dose using the formalism presented in equation 2:

$$D(cGy) = \frac{(D/\text{\# of incident particles})_{MC\ individual\ simulaiton}}{(D/\text{\# of incident particles})_{MC\ calibration\ simulaiton}} \times \frac{1\ cGy}{MU} \times MU_{del} \quad (2)$$

where $MU_{del}$ is the monitor units (MU) delivered by a linear accelerator. In this formula $\frac{D}{\text{\# of incident particles}}$ represents the dose scored per number of incident particles in a Monte Carlo simulation. The calibration simulation was performed in water for a square field of $10 \times 10$ cm$^2$ and SSD of 100 cm and dose was scored at a depth of 10 cm.

Delivery log files were converted into input files for MC simulations using an in-house Python script. The photon and electron cut-off energies (PCUT and ECUT) were set to 0.01 and 0.7 MeV and the electron range rejection was set to 2 MeV for all simulations. The target mean relative statistical uncertainty (Chetty *et al* 2006) for these simulations was 0.4% which was calculated over all voxels with doses greater than 50% of the maximum dose. Achieving this statistical uncertainty required simulating $1.5 \times 10^8$ and $3.0 \times 10^8$ histories on the stationary and deforming geometries, respectively, with approximate CPU core hours equal to 12–15 and 31–35 per calculations. All simulations were performed on the Carleton University Physics Research Compute Cluster which consists of 644 processing cores Intel Xeon CPUs at 2.50–3.00 GHz frequency.

*2.5.2. Dose calculations: Stationary and deforming phantom*

The 3DCT scans of the EOI phantom state were resampled to a resolution of $0.05 \times 0.05 \times 0.2$ cm$^3$ to generate the dose calculation geometry. Assignment of voxel densities followed the approach introduced by Seco (Seco and Evans 2006) that enables direct calculation of $D_w$ in MC simulations to be compared against film and RADPOS measurements.

Source 21 (Lobo and Popescu 2010) of DOSXYZnrc was used to simulate the BEAMnrc linac model for the simulations performed in this work. 4DdefDOSXYZnrc simulations utilized the DVF exported from Velocity along with the respiratory motion trace recorded with RADPOS (tumour center) during irradiations to model the phantom motion and deformation. For each particle incident on the phantom, the magnitude and direction of deformations of the reference geometry are determined by scaling the DVF by the magnitude of the respiratory motion at the appropriate time point (respiratory phase). In order to conserve mass between the reference and deformed geometries, the densities of the deformed voxels are recalculated. The incident particle is transported through the deformed geometry and its energy deposition is scored. No mapping of dose between geometries is required since the same dose calculation grid is retained between the reference and deformed geometries. The cumulative dose in each voxel is calculated by dividing the total energy deposition by the mass of the reference voxel (Gholampourkashi *et al* 2017, 2018b).

*2.6. Comparison metrics*

Dose profiles along the motion path of the phantom (S-I direction) from MC simulations were compared against film measurements. Measurement of the profiles was limited by the deformations in the superior side of the phantom where the piston is placed. Deformations of the phantom introduced a constraint on using wider film pieces to acquire 2D dose distribution. To evaluate dose profiles from simulations, a 1D gamma analysis (Low *et al* 1998) with a 2% dose-difference and 2 mm distance-to-agreement criterion with film as the reference was utilized. The dose threshold used for gamma analysis was set to be 5% of the evaluated maximum dose. To match the dose grid resolution of film with the one from MC simulations (2 mm), a moving average filter was applied to the film readings. Also, point dose comparisons between MC simulations and measurements at the center of the tumor (film and RADPOS) as well as top and bottom surfaces of the plug (RADPOS) are quoted as percentage of the measured dose value at the point of measurement.

## 3. Results

*3.1. Assessment of motion reproducibility*

Measured motion magnitudes and their reproducibility in S-I, A-P and L-R directions for various diaphragm amplitudes for sinusoidal motion are shown in Figure 6.

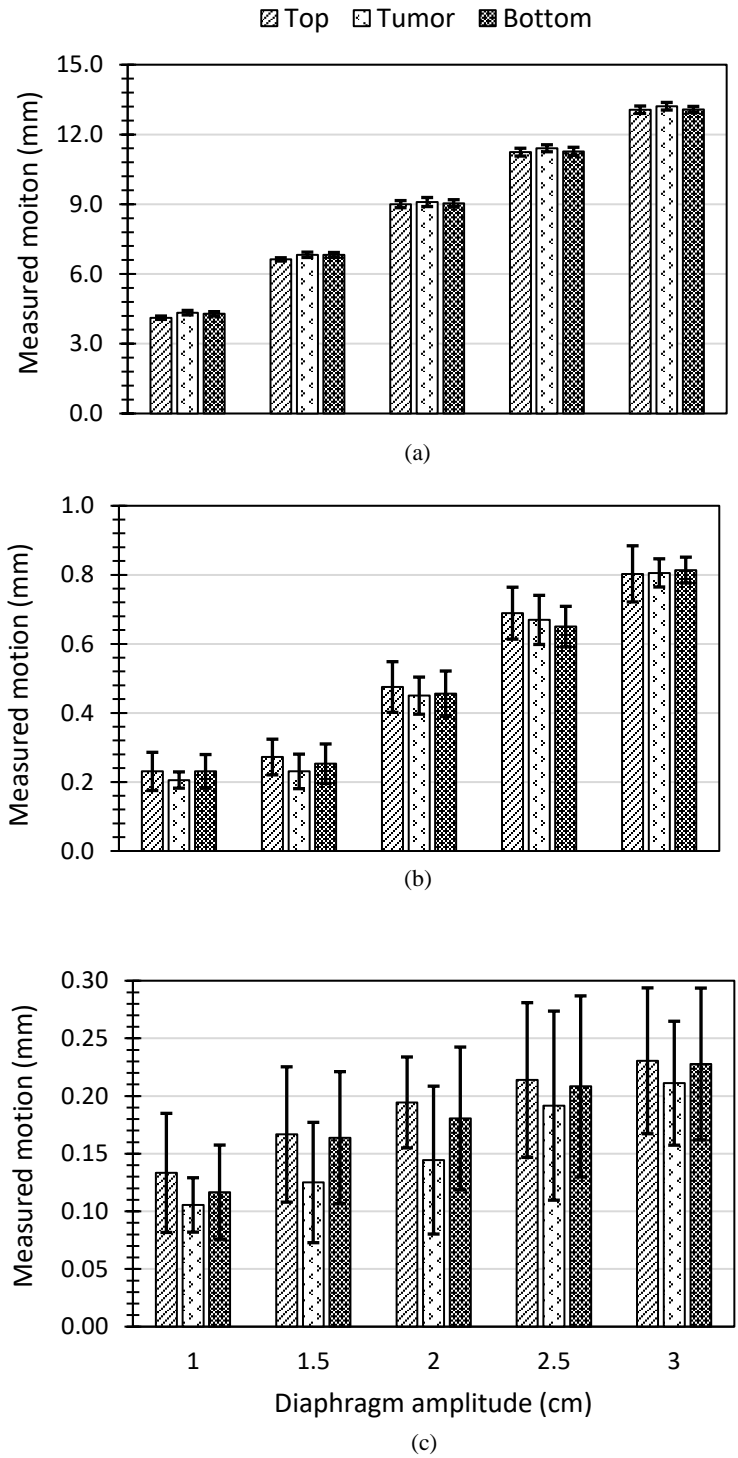

**Figure 6.** Average range of motion in (a) S-I, (b) A-P and (c) L-R directions measured by RADPOS detectors at the tumor, top and bottom plug surfaces for sinusoidal motion with various diaphragm amplitudes. Error bars show reproducibility of the motion (combined standard deviations for one amplitude is the square root of the sum of the square of individual standard deviations).

The motion amplitude did not show significant variations to changes in the period of the sinusoidal motion. The variations were observed to be less than 0.1 mm in S-I and 0.02 mm in A-P and L-R directions. The overall reproducibility of the P-P amplitudes was measured to be within 0.2, 0.1 and 0.1 mm for all motions in S-I, A-P and L-R directions, respectively. The S-I motion measured by all three RADPOS detectors was very similar (within 0.2 mm or 4%) while differences of almost 0.04 mm (~ 20%) and 0.05 mm (~ 30%) were observed in the motion measured in A-P and L-R directions, respectively.

Maximum intra- and inter-day variations for S-I, A-P and L-R motion amplitude for all three RADPOS detectors are shown in Table 1. For the inter-day measurements, the RADPOS probes were not removed between consecutive measurements.

**Table 1.** Inter- and intra-day amplitude variations in S-I, A-P and L-R directions for tumor, top and bottom RADPOS detectors for sinusoidal motion profiles.

|           | Measurement point | A-P (cm) | L-R (cm) | S-I (cm) |
|-----------|-------------------|----------|----------|----------|
| Inter-day | Tumor Center      | 0.02     | 0.02     | 0.09     |
|           | Top surface       | 0.03     | 0.02     | 0.08     |
|           | Bottom surface    | 0.02     | 0.02     | 0.07     |
| Intra-day | Tumor Center      | 0.01     | 0.01     | 0.02     |
|           | Top surface       | 0.02     | 0.01     | 0.03     |
|           | Bottom surface    | 0.02     | 0.02     | 0.02     |

For the irregular respiratory motion profiles the motion amplitude varies within the motion profile. For the motion profiles shown in Figure 6, these amplitudes were calculated to be 9.78 ± 1.19, 6.56 ± 1.90 and 13.1 8 ± 0.18 mm, respectively. Reproducibility of the average P-P amplitudes as well as the amplitude variations in 3D from all simulated profiles are shown in Table 2 for the three RADPOS detectors.

**Table 2.** Reproducibility of the average P-P amplitude and amplitude variations in S-I, A-P and L-R directions for tumor, top and bottom RADPOS detectors for irregular motion profiles. Values represent the standard deviation of the average P-P amplitude and average value of amplitude variations across individual motion traces, respectively.

|  | Measurement point | A-P (cm) | L-R (cm) | S-I (cm) |
|---|---|---|---|---|
| Average P-P amplitude | Tumor Center | 0.025 | 0.012 | 0.038 |
|  | Top surface | 0.023 | 0.012 | 0.039 |
|  | Bottom surface | 0.022 | 0.010 | 0.037 |
| P-P amplitude variations | Tumor Center | 0.004 | 0.004 | 0.015 |
|  | Top surface | 0.005 | 0.004 | 0.012 |
|  | Bottom surface | 0.003 | 0.006 | 0.014 |

Maximum inter-day variations of the amplitude are shown in table 3 for S-I, A--P and L-R directions for all three RADPOS detectors.

**Table 3.** Inter-day amplitude variations in S-I, A-P and L-R directions for tumor, top and bottom RADPOS detectors for irregular motion profiles.

|  | Measurement point | A-P (cm) | L-R (cm) | S-I (cm) |
|---|---|---|---|---|
| Average P-P amplitude | Tumor Center | 0.027 | 0.028 | 0.087 |
|  | Top surface | 0.031 | 0.028 | 0.094 |
|  | Bottom surface | 0.024 | 0.025 | 0.093 |
| P-P amplitude variations | Tumor Center | 0.010 | 0.013 | 0.030 |
|  | Top surface | 0.013 | 0.014 | 0.030 |
|  | Bottom surface | 0.008 | 0.014 | 0.036 |

One of the applications of RADPOS as a real-time motion detector is to enable examination of the correlation between the motion of the diaphragm to the motion of any point of interest (e.g. tumor) inside the phantom. Figure 7 shows such correlations for sinusoidal (periods of 2 and 7 s) and irregular motion profiles shown in Figure 4(a and c) in the form of hysteresis plots for S-I direction.

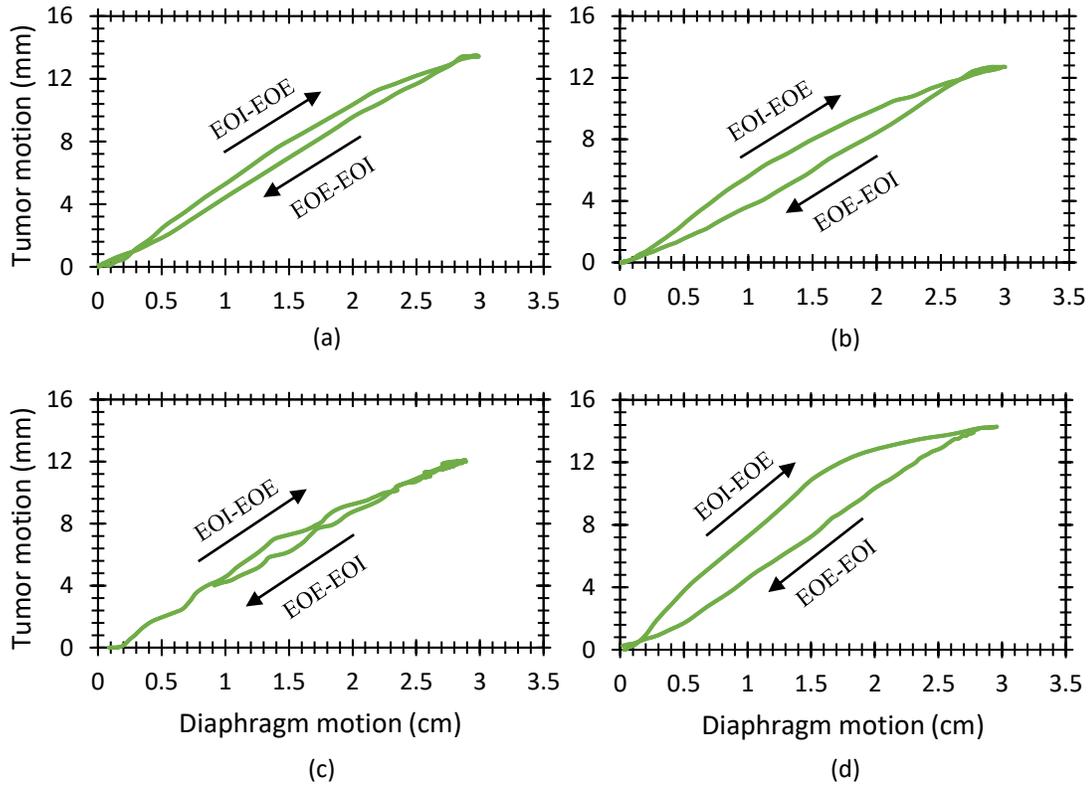

**Figure 7.** Correlation curves between motion of the tumor and diaphragm for P-P diaphragm amplitude of 3 cm. Correlations are shown in the form of hysteresis plots for sinusoidal motion profiles with periods of (a) 4 s, (b) 7 s as well as (c) typical (Figure 4(a)) and (d) large hysteresis (Figure 4(c)) respiratory motion profiles.

## 3.2. Validation of image registration

Performance of the deformable image registration algorithm was evaluated both visually (Figure 8) and quantitatively (Table 4). Mean values of registration errors with their standard deviations are shown in Table 4 for A-P, L-R and S-I direction. The overall 3D registration error is presented in the Table as well. The 3D registration error for each landmark and the three RADPOS detectors is calculated by adding errors in all three directions in quadrature.

**Table 4.** Mean registration errors with their standard deviations in A-P, L-R, S-I and 3D.

| Direction of motion | Mean registration error (cm) |
|:---:|:---:|
| A-P | 0.05 ± 0.03 |
| L-R | 0.04 ± 0.03 |
| S-I | 0.08 ± 0.05     |
| 3D  | 0.12 ± 0.04     |

The maximum and minimum 3D error values were measured to be 0.18 and 0.04 cm, respectively. The limiting factor in achieving a better registration accuracy was the resolution of the CT image slices which was 0.2 cm.

Figures 8(a) and 8(b) show the overlaid EOI and EOE images before and after registration. The difference map shown in Figure 8(c) presents the differences between the deformed EOE image from the EOI. Regions in light gray present the lowest differences between the deformed EOE and EOI images. In Figure 8(d) the Jacobian map of the registration is representative of how voxels change in size once deformations are applied. In this colormap, green regions represent no volume change between deformed EOE and EOI images. Blue and red regions present shrinkage (reduced voxel size) and growth (enlarged voxel size).

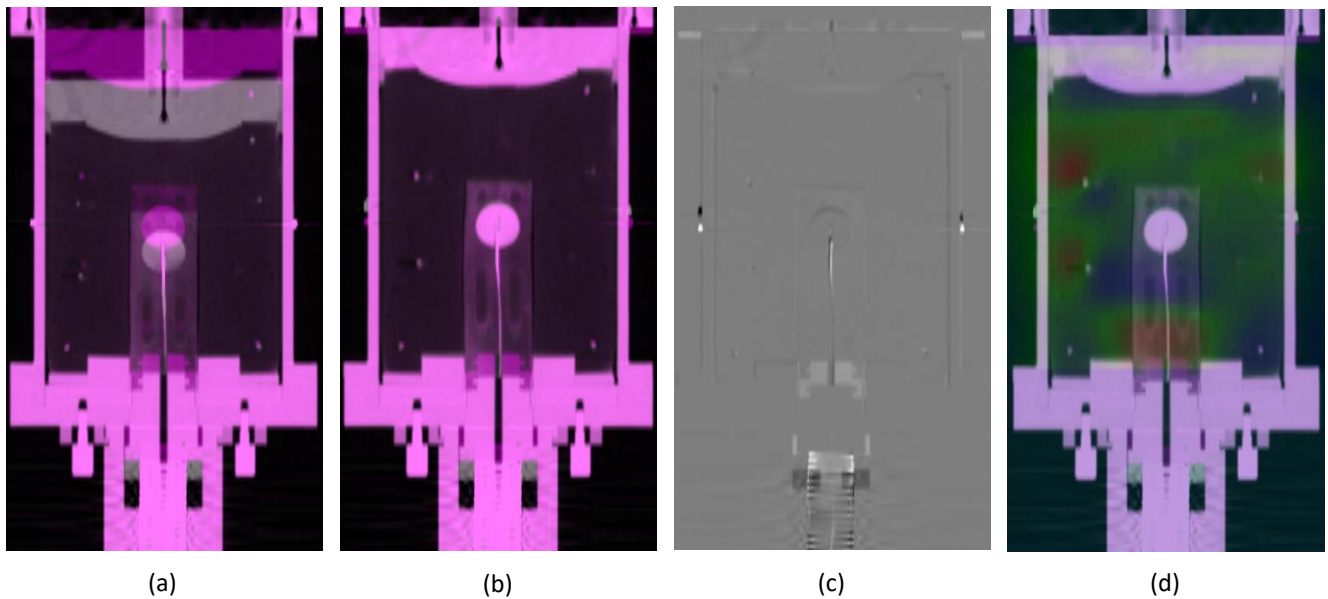

(a)  (b)  (c)  (d)

**Figure 8.** Visual evaluation of the deformable image registration on coronal view: (a) non-deformed EOE (gray) overlaid on EOI (pink), (b) deformed EOE overlaid on EOI, (c) deformed EOE subtracted from EOI and (d) Jacobian of the registration.

### 3.3. Dosimetric comparisons

Figure 9 shows sample dose profiles from MC simulations and film measurements for the respiratory traces previously shown in Figure 4. Corresponding 1D gamma passing rates (2%/2 mm) are shown in Table 5. In addition, average values of gamma passing rates for all sets of irradiations were calculated and are shown in this table.

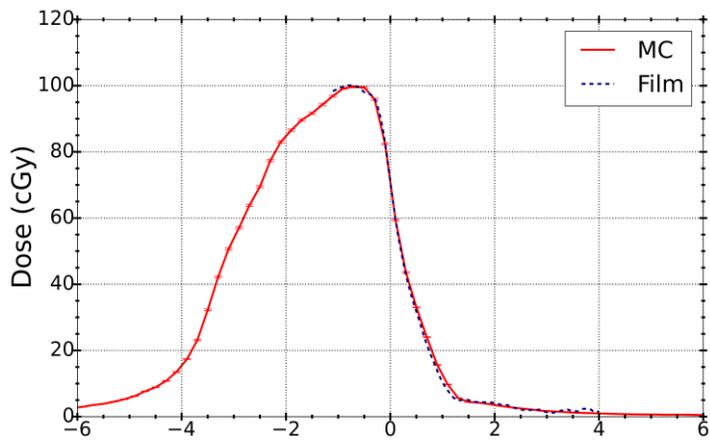
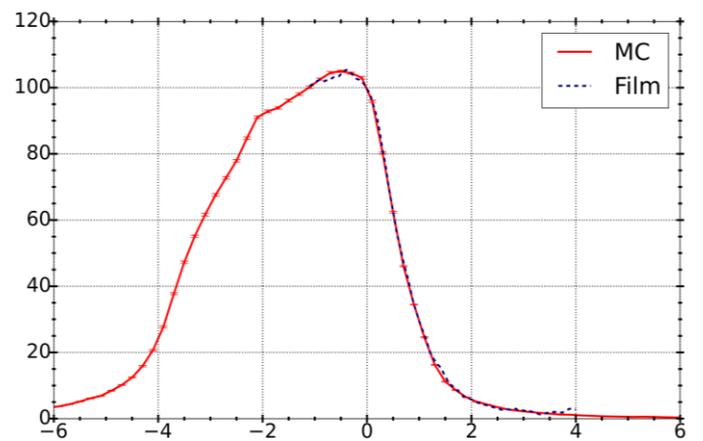
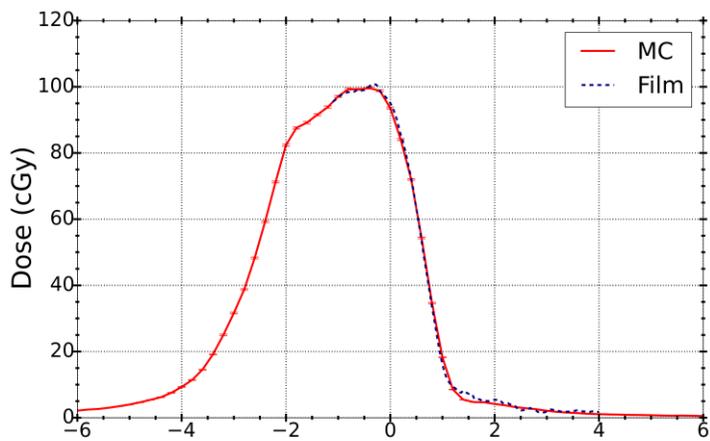
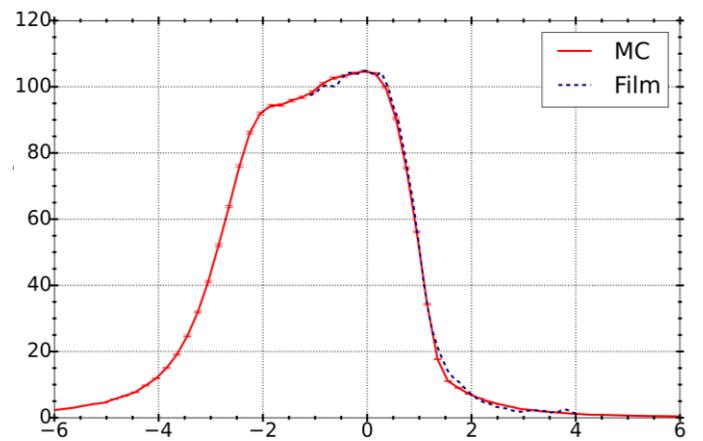
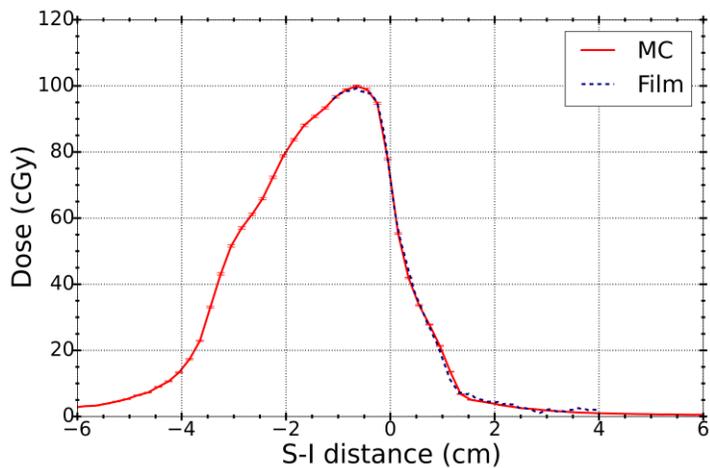
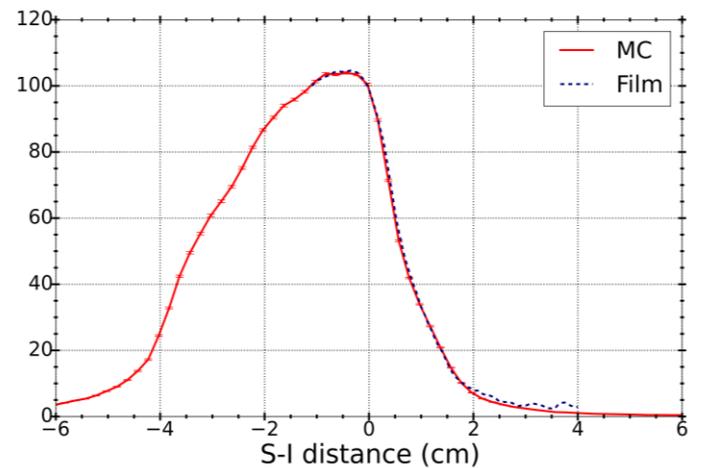

**Figure 9.** Comparison of dose profiles for 3×3 cm$^2$ (left) and VMAT (right) beam deliveries on the breathing deformable phantom along the S-I direction, for the typical (Figure 4(a)) (top row), highly irregular (Figure 4(b)) (middle row) breathing motions as well the breathing motion with large hysteresis (Figure 4(c)) (bottom row).

**Table 5.** Passing rates of 1D gamma comparisons of 2%/2 mm criteria for MC simulations against film measurements on the breathing deformable phantom for typical, highly irregular and large hysteresis respiratory motion profiles. First column of gamma passing rates corresponds to the dose profiles shown in Figure 9. The average values from all irradiation sets are shown in the second column.

| Plan | Motion | 2%/2 mm 1D gamma passing rate (%) | |
| --- | --- | --- | --- |
| | | Single profile | Average of 3 measurements |
| Static 3×3 cm$^2$ | Typical (Figure 9(top-left)) | 100.0 | 97.7 ± 2.9 |
| | Highly irregular (Figure 9(middle-left)) | 98.9 | 98.0 ± 2.6 |
| | Large hysteresis (Figure 9(bottom-left)) | 100.0 | 99.3 ± 1.2 |
| VMAT | Typical (Figure 9(top-right)) | 99.7 | 97.8 ± 1.9 |
| | Highly irregular (Figure 9(middle-right)) | 98.9 | 98.1 ± 1.3 |
| | Large hysteresis (Figure 9 (bottom-right)) | 100.0 | 97.9 ± 1.8 |

Also, dose values measured and simulated at the center of tumor are shown in Tables 6 with their corresponding statistical and experimental uncertainties.

**Table 6.** Dose values at the center of tumor from measurements with film and RADPOS as well as MC simulations on the breathing deformable phantom during typical, highly irregular and large hysteresis respiratory motion profiles.

| Plan | Motion | Irradiation # | Dose (cGy) | | |
|---|---|---|---|---|---|
| | | | | Measurements | |
| | | | MC | Film | RADPOS |
| Static 3×3 cm$^2$ | Large hysteresis | - | 72.1 ± 0.4% | 71.8 ± 2.3% | 73.5 ± 2.4% |
| | Typical | 1 | 70.9 ± 0.4% | 70.9 ± 2.3% | 70.9 ± 2.4% |
| | | 2 | 77.3 ± 0.4% | 78.7 ± 2.3% | 77.8 ± 2.4% |
| | | 3 | 79.8 ± 0.4% | 79.5 ± 2.3% | 79.4 ± 2.4% |
| | Highly irregular | 1 | 97.4 ± 0.4% | 98.6 ± 2.3% | 97.2 ± 2.4% |
| | | 2 | 93.2 ± 0.4% | 95.0 ± 2.3% | 95.1 ± 2.4% |
| | | 3 | 88.6 ± 0.4% | 90.4 ± 2.3% | 91.1 ± 2.4% |
| VMAT | Large hysteresis | - | 98.8 ± 0.4% | 98.9 ± 2.3% | 98.0 ± 2.4% |
| | Typical | 1 | 98.8 ± 0.4% | 99.3 ± 2.3% | 100.0 ± 2.4% |
| | | 2 | 99.4 ± 0.4% | 99.6 ± 2.3% | 100.5 ± 2.4% |
| | | 3 | 97.8 ± 0.4% | 97.3 ± 2.3% | 100.3 ± 2.4% |
| | Highly irregular | 1 | 104.3 ± 0.4% | 104.5 ± 2.3% | 104.3 ± 2.4% |
| | | 2 | 102.0 ± 0.4% | 102.6 ± 2.3% | 101.2 ± 2.4% |
| | | 3 | 101.0 ± 0.4% | 103.0 ± 2.3% | 100.0 ± 2.4% |

From the values in Table 6 we can see that for the majority of the irradiations, the measured and simulated dose values at the center of tumor agree within 2% of each other. Exceptions are the third beam deliveries of the static 3×3 cm² plan for the highly irregular motion and the VMAT plan for the typical respiratory motion that show agreement within 3% between MC simulations and RADPOS measurements which still lies within the 2σ of the experimental uncertainties (4.8%). Overall, for the irradiations done for the respiratory motion with large hysteresis trace, the same level of agreement (i.e. better than 3%) was found to be true. In addition, from dose profiles shown in Figure 9, we can see that sharp dose gradients exist in the delivered dose due to respiratory motion. This is especially prominent for the 3×3 cm² plan which adds an intrinsic dose gradient compared to the VMAT plan. For some motion traces (typical and large hysteresis), the total positional/reading uncertainties of dose values can be as high as 8% for the 3×3 cm² plan which is observable from the level of dose gradient in profiles in Figures 9(top-left) and 9(bottom-left). These uncertainties are calculated incorporating the percentage point dose differences within a 2 mm voxel including positional uncertainties in 3D (Left-Right, Sup-Inf and Ant-Post). Considering that the largest component of the motion happens in the S-I direction, such uncertainties could be twice as the overall values if we take only S-I direction into consideration. For the highly irregular motion trace on the other hand, these uncertainties did not exceed 3-4%. Table 7 presents dose values measured and simulated at the bottom surface of the plug with their corresponding experimental and statistical uncertainties.

**Table 7.** Dose values at the bottom surface of the plug from MC simulations and RADPOS measurements on the breathing deformable phantom during typical, highly irregular and large hysteresis respiratory motion profiles.

| Plan | Motion | Irradiation # | Dose (cGy) Bottom surface | |
|---|---|---|---|---|
| | | | MC | RADPOS |
| Static 3×3 cm² | Large hysteresis | - | 55.1 ± 0.4% | 53.5 ± 2.4% |
| | Typical | 1 | 51.9 ± 0.4% | 51.6 ± 2.4% |
| | | 2 | 56.6 ± 0.4% | 55.0 ± 2.4% |
| | | 3 | 58.2 ± 0.4% | 58.3 ± 2.4% |
| | Highly irregular | 1 | 77.2 ± 0.4% | 77.9 ± 2.4% |
| | | 2 | 75.3 ± 0.4% | 77.3 ± 2.4% |
| | | 3 | 71.3 ± 0.4% | 73.6 ± 2.4% |
| VMAT | Large hysteresis | - | 65.7 ± 0.4% | 63.5 ± 2.4% |

|  |  |  |  |
|---|---|---|---|
| Typical | 1 | 66.0 ± 0.4% | 65.8 ± 2.4% |
|  | 2 | 65.5 ± 0.4% | 66.8 ± 2.4% |
|  | 3 | 65.8 ± 0.4% | 63.3 ± 2.4% |
| Highly irregular | 1 | 82.0 ± 0.4% | 80.6 ± 2.4% |
|  | 2 | 79.0 ± 0.4% | 78.0 ± 2.4% |
|  | 3 | 77.4 ± 0.4% | 76.2 ± 2.4% |

At the bottom surface of the plug, measured and simulated dose values shown in Table 7 show a better than 3% level of agreement. Similar to the tumor dose values, exceptions are the two same irradiations which do not show an agreement larger than 4% which is within the 2σ of the experimental uncertainties (4.8%) as well. The total level of agreement for all irradiations for the respiratory motion with the large hysteresis trace was observed to be not larger than 4%. Once studied, approximately same level of dose gradient and as a result dose reading/positional uncertainty as for the center of tumor was observed at this dose point.

Measured and simulated dose values as well as their experimental and statistical uncertainties at the top surface of the plug are shown in Table 8. An agreement of better than 5% is observable for those values shown in this table. This level of agreement complies with the fact that this dose point is initially positioned in a high dose gradient region. It should be noted that since this point dose is placed close to the edge of the beam, the motion results in a larger drop in the dose value compared to dose values at the tumor center and the bottom surface. The overall dose reading/positional uncertainties are approximately 12% and 10% for the 3×3 cm$^2$ and VMAT plan deliveries while considering only the S-I direction these values can go as high as 20%. In average, same level of dose difference were observed for all irradiations with the large hysteresis respiratory motion profile with some dose differences slightly larger than 5%.

In order to explain the dose differences observed between three irradiations (beam-on at respectively 0, 40 and 90 s after the motion starts) for the typical and highly irregular motion profiles, respiratory traces recorded with RDPOS during beam deliveries were extracted. The fraction of time that was spent in each respiratory phase is shown in Figures 10 and 11 for these two traces, respectively, for both plan deliveries.

**Table 8.** Dose values at the top surface of the plug from MC simulations and RADPOS measurements on the breathing deformable phantom during typical, highly irregular and large hysteresis respiratory motion profiles. This RADPOS was placed at an offset position of 1 cm from the tumor center.

| Plan | Motion | Irradiation # | Dose (cGy) Top surface | |
|---|---|---|---|---|
| | | | MC | RADPOS |
| | Large hysteresis | - | 18.0 ± 0.4% | 18.5 ± 2.2% |
| | Typical | 1 | 13.8 ± 0.4% | 13.2 ± 2.2% |
| | | 2 | 15.4 ± 0.4% | 15.0 ± 2.2% |
| | | 3 | 16.4 ± 0.4% | 15.9 ± 2.2% |
| | Highly irregular | 1 | 17.1 ± 0.4% | 17.9 ± 2.2% |
| | | 2 | 17.4 ± 0.4% | 18.2 ± 2.2% |
| | | 3 | 14.8 ± 0.4% | 15.1 ± 2.2% |
| | Large hysteresis | - | 22.9 ± 0.4% | 23.6 ± 2.2% |
| | Typical | 1 | 21.8 ± 0.4% | 22.6 ± 2.2% |
| | | 2 | 23.8 ± 0.4% | 24.0 ± 2.2% |
| | | 3 | 22.3 ± 0.4% | 23.4 ± 2.2% |
| | Highly irregular | 1 | 33.2 ± 0.4% | 34.1 ± 2.2% |
| | | 2 | 31.8 ± 0.4% | 32.5 ± 2.2% |
| | | 3 | 32.9 ± 0.4% | 33.8 ± 2.2% |

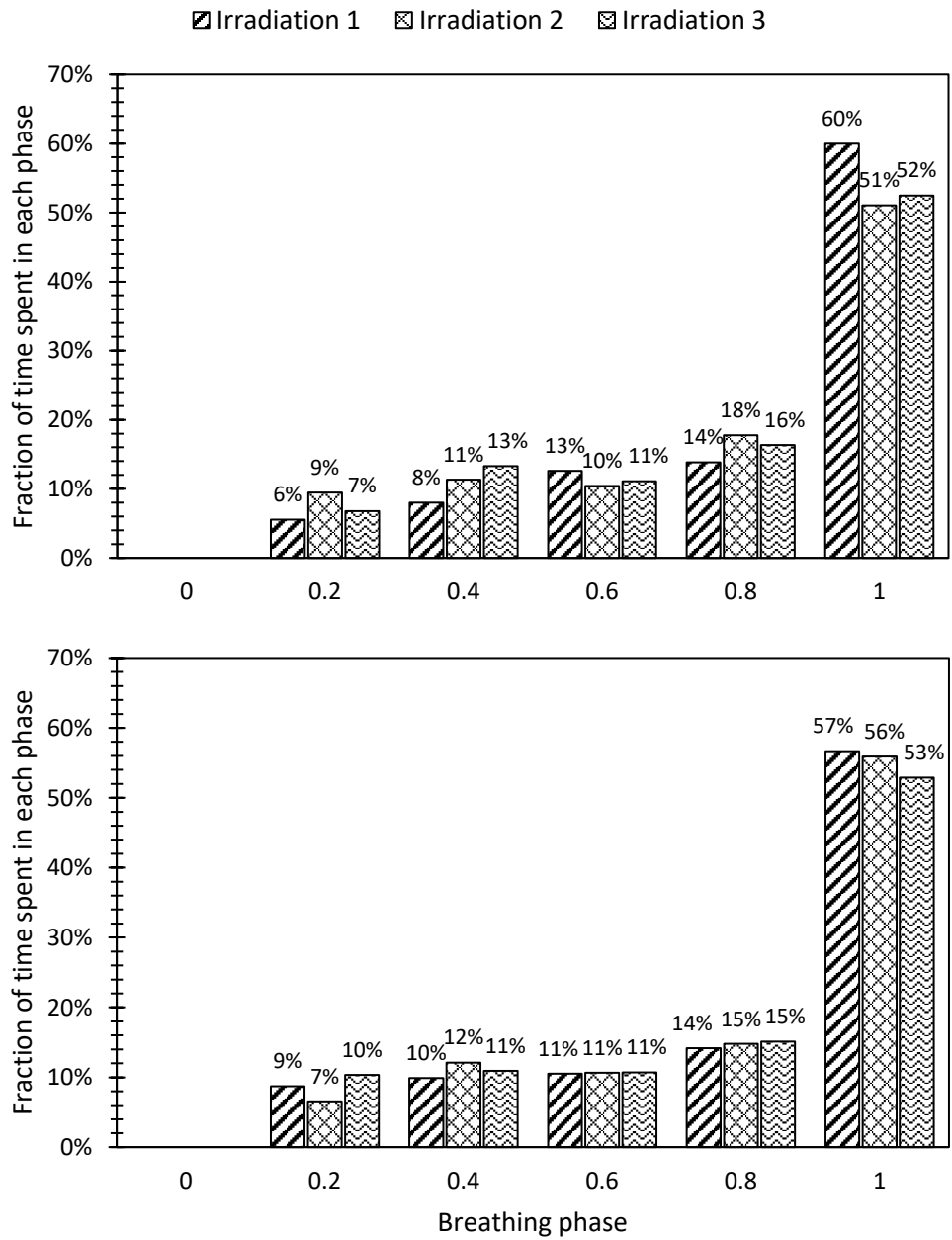

**Figure 10.** Fraction of the time spent in each breathing phase during 3×3 cm² (top) and VMAT (bottom) beam deliveries on the deforming phantom for three irradiations with the typical respiratory motion profile

From the values shown in Figure 10 we can see that 60% of the first static 3×3 cm$^2$ beam delivery happened while the phantom was in the EOE (fully compressed or breathing phase 1) state, whereas for the other two beam deliveries this value reduces to approximately 50%. When studied in more details, it was observed that 45% of the total time spent in EOE for the first irradiation was spent in this state without any transitions to EOI. This explains the lower dose values in the first irradiation compared the two other irradiations. As for VMAT deliveries, it can be seen that almost equal times are spent in each phase for the three irradiations since VMAT plans take longer time compared to static plans (~ 67 s vs ~15 s) to deliver.

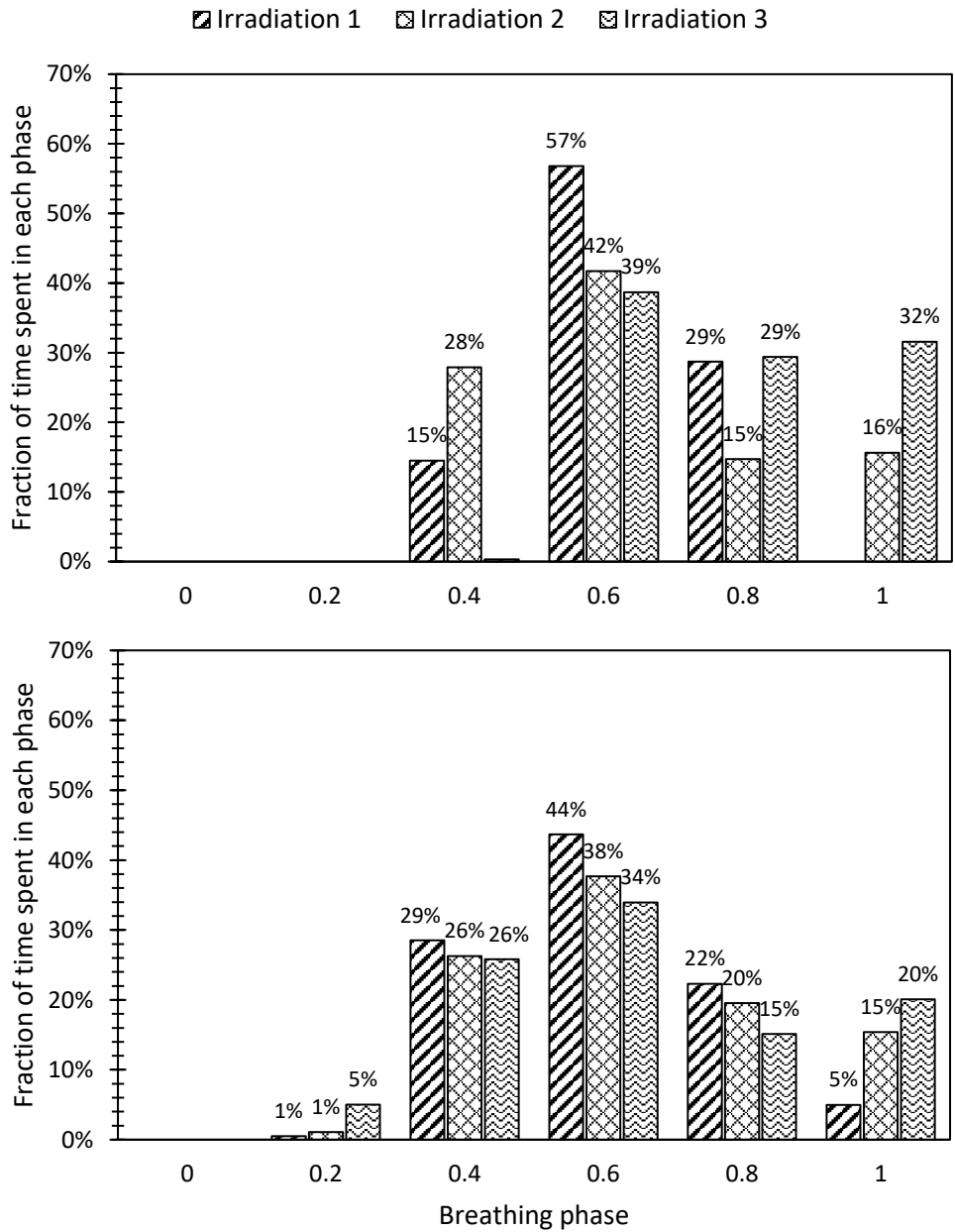

**Figure 11**. Fraction of the time spent in each breathing phase during 3×3 cm$^2$ (top) and VMAT (bottom) beam deliveries on the deforming phantom for three irradiations with the highly irregular respiratory motion profile shown in Figure 4(b). Breathing phase is the RADPOS normalized displacement that was recorded during beam deliveries.

For the static 3×3 cm$^2$ beam delivery of the highly irregular motion profile, it can be seen in top plot of Figure 11 that over 70% of the beam delivery was done while the phantom was in the early- and mid-inhale states. However, looking the last irradiation, only 40% of the beam was delivered at this state and the remaining 60% was delivered as phantom was in the mid-exhale and EOE states. As a result, the dose delivered during this beam delivery was lower compared to the first two deliveries. On the second plot related to the VMAT delivery it can be seen that approximately 75% of the first irradiation happened while the phantom was in the early- and mid-inhale states which is almost 10-15% higher than similar values for the two last irradiations. As a result, we can see some dose differences between the deliveries in this case.

## 4. Discussion

In this study 4D Monte Carlo simulations using the 4DdefDOSXYZnrc user code to calculate the dose delivered to a breathing deformable lung phantom during static and VMAT beam deliveries while phantom moved with irregular breathing motions were validated. Film was placed inside the lung inserts of the phantom to measure dose distribution along the motion paths (S-I). Point dose measurements with RADPOS were performed inside the GTV (i.e. tumor). Point doses were also measured outside the tumor (still inside the lung). Recorded displacements with RADPOS inside the GTV as well as DVFs generated by DIR were used as input to 4DMC simulations to model the motions/deformations of phantoms.

Our results (Figure 9, Tables 5-8) showed that the overall agreement of point dose values at the center of the tumor from MC simulations and measurements by film were within 2% of each other. Dose differences between RADPOS measurements and MC simulations did not exceed 3% which is not larger than 2σ of the experimental uncertainties with RADPOS measurements. As for the dose points outside the tumor (i.e. top and bottom surfaces of the plug), simulations and measurements were found to have an average agreement of 6% or better. These agreements were found to be within the calculated positional uncertainties of these dose points. The agreements between the simulated and measured dose profiles (along the S-I direction), were good as well. Gamma comparisons of 2%/2 mm showed an overall passing rate of almost 94% or better.

In order to investigate the impact of starting phase of the respiratory cycle, treatments for deliveries of both plans were started 0, 40 and 90 s after the deformable phantom started to move with typical and highly irregular respiratory profiles. Different doses can be measured depending on the amount of time a target volume spends in EOI, EOE or transitions between these two states. In this work it was observed that for static plan deliveries (Figures 10&11 (top)), the dose delivered to the tumor as well as the bottom surface of the plug could change by 10-12% once they spent almost 45% of the delivery time in the EOE without any transitions to EOI. Similar fractions of time spent in EOE resulted in a dose change of 15-20% at the dose point on the top surface of the tumor due to the fact that it was in the penumbra region of the beam. For VMAT deliveries (Figures 10&11 (bottom)), the longer treatment times compared to static deliveries help reduce the impact of differences in the respiratory cycle during several deliveries of the same treatment plan. As a result, dose differences between deliveries may not be as large as seen for static treatment plans. In the case of VMAT deliveries for the typical breathing trace where the target volume spends almost equal amounts of time in the EOE and EOI during different deliveries, dose differences of less than 1% were observed. On the other hand, with highly irregular respiratory traces the case could be different. Differences of almost 3-4% were observed for this respiratory profile between the one VMAT beam delivery with 10-15% more time spent in early-, mid-inhale compared to the other two deliveries. These results imply the importance of accurate detection of the starting phase of the breathing cycle and how it may impact the 4D dose calculations. In this work, uncertainty in PC clock synchronization of RADPOS and linac as well as system delays caused by the temporal resolution of RADPOS (100 ms) are two main sources that can affect proper detection of the start of the breathing trace.

Comparison of dose values from the static beam deliveries from our previous study (Gholampourkashi *et al* 2018b) for the sinusoidal respiratory motion and current study with respiratory motion with large hysteresis traces revealed approximate decreases of 15% of the dose delivered to GTV (i.e. center of tumor). This result was expected considering the time spent in each breathing phase during beam deliveries for each of these motion profiles. While moving the sinusoidal motion, these dose points spend almost equal amounts of times from EOI-EOE and EOE-EOI. However, these times change to 20% and 80% while the phantom moves with the respiratory motion trace with the large hysteresis.

A current limitation of our 4DMC tool is that it relies on deformable registration between only the two extreme respiratory phases (i.e. EOE and EOI) to model the anatomical motion. Although this was found to be adequate for modeling the deformations of our phantom, it will not be able to model the hysteresis (Suh *et al* 2008) that can occur in patient respiratory motion. We are working to extend the motion model to include deformation vectors determined from registrations between multiple respiratory phases of a 4DCT dataset with aim of applying it to patient 4D dose reconstruction.

## 5. Conclusions

We investigated and established the accuracy of 4D Monte Carlo simulations, using the EGSnrc/4DdefDOSXYZnrc user code, of dose delivered to a programmable deforming phantom in presence of realistic breathing motion. Measurements were performed on an Elekta Infinity linac equipped with Agility MLC during static square and VMAT plan deliveries. Delivery log files were used to reproduce the measurements during these deliveries. Our findings demonstrate that combining the motion recorded by RADPOS with DVFs generated by a reliable DIR algorithm into our 4DMC simulations, leads to accurate calculation of cumulative dose delivered to a deforming anatomy undergoing irregular breathing motion.